\documentclass{llncs}
\usepackage{wrapfig}
\usepackage[ruled,linesnumbered,noend]{algorithm2e}
\usepackage{citesort}
\usepackage{multicol}
\usepackage{amssymb}
\usepackage{amsmath,bbm}
\usepackage{latexsym,epic,eepic}
\usepackage{complexity}
\usepackage{multirow}
\usepackage{algorithmic}
\usepackage{url}
\usepackage{times}
\usepackage{tikz}
\usetikzlibrary{shapes}
\usetikzlibrary{automata,positioning}

\newcommand{\OMIT}[1]{}

\newcommand{\defn}[1]{\textit{#1}} 
\newcommand{\mcl}[1]{\ensuremath{\mathcal{#1}}}

\newcommand{\struct}{\mathfrak{S}}

\newcommand{\Trans}[1]{\struct_{#1}}
\newcommand{\transysDom}{S}
\newcommand{\transys}{\ensuremath{\langle \transysDom; \to\rangle}}

\newcommand{\PAD}{\ensuremath{\mcl{PAD}}}
\newcommand{\divides}{|}
\newcommand{\PA}{\ensuremath{\mcl{P}}}
\newcommand{\ialphabet}{\ensuremath{A}}

\newcommand{\Var}{\ensuremath{\mcl{V}}}

\newcommand{\N}{\ensuremath{\mathbb{N}}}

\newcommand{\Z}{\ensuremath{\mathbb{Z}}}

\newcommand{\vecX}{\ensuremath{\text{\rm \bfseries x}}}
\newcommand{\vecY}{\ensuremath{\text{\rm \bfseries y}}}
\newcommand{\vecZ}{\ensuremath{\text{\rm \bfseries z}}}

\newcommand{\vecW}{\ensuremath{\text{\rm \bfseries w}}}
\newcommand{\vecV}{\ensuremath{\text{\rm \bfseries v}}}

\newcommand{\problemx}[3]{
\par\noindent\underline{\sc#1}\par\nobreak\vskip.2\baselineskip
\begingroup\clubpenalty10000\widowpenalty10000
\setbox0\hbox{\bf Instance: }\setbox1\hbox{\bf Question: }
\dimen0=\wd0\ifnum\wd1>\dimen0\dimen0=\wd1\fi
\vskip-\parskip\noindent
\hbox to\dimen0{\box0\hfil}\hangindent\dimen0\hangafter1\ignorespaces#2\par
\vskip-\parskip\noindent
\hbox to\dimen0{\box1\hfil}\hangindent\dimen0\hangafter1\ignorespaces#3\par
\endgroup}

\newcommand{\Fu}{\text{\bf F}}
\newcommand{\Gl}{\text{\bf G}}

\newcommand{\CA}{\ensuremath{\mathcal{C}}} 
\newcommand{\Aut}{\ensuremath{\mathcal{A}}} 
\newcommand{\transrel}{\ensuremath{\Delta}} 

\newcommand{\controls}{\ensuremath{Q}} 
\newcommand{\Lang}{\ensuremath{\mathcal{L}}} 
\newcommand{\AutRun}{\ensuremath{\rho}} 

\newcommand{\To}{\Rightarrow}

\newcommand{\counters}{\ensuremath{X}}
\newcommand{\ctr}{\ensuremath{x}}

\newcommand{\ID}{\ensuremath{\text{\texttt{ID}}}}
\newcommand{\SUB}{\ensuremath{\text{\texttt{SUB}}}}
\newcommand{\DEC}{\ensuremath{\text{\texttt{DEC}}}}
\newcommand{\LEN}{\ensuremath{\text{\textsc{Len}}}}

\newcommand{\geqReg}{\ensuremath{\trianglerighteq}} 
\newcommand{\leqReg}{\ensuremath{\trianglelefteq}} 
\newcommand{\gtReg}{\ensuremath{\triangleright}}

\newcommand{\ACCELERATE}{\ensuremath{\text{\textsc{L-Accelerate}}}}

\def\soln{{\sigma}}
\def\Vars{{V}}
\def\set#1{{\{ #1 \}}}

\SetCommentSty{mycommfont}
\newlength{\commentWidth}
\newif\ifdraft\draftfalse
\ifdraft
\newcommand\anthony[1]{{\color{blue}
[#1 - \textbf{Anthony}]}}
\newcommand\rupak[1]{{\color{magenta}
[#1 - \textbf{Rupak}]}}

\newcommand\todo[1]{}
\else
\newcommand\anthony[1]{}
\newcommand\rupak[1]{}

\newcommand\todo[1]{}
\fi

\newcommand\shortlong[2]{#2}

\usepackage{color}
\shortlong{}{\pagestyle{plain}}

\title{
Quadratic Word Equations with Length Constraints, Counter Systems, and Presburger Arithmetic with Divisibility 
}
\author{ 
    Anthony W. Lin\inst{1} \and
    Rupak Majumdar\inst{2} 
}
\institute{
    Oxford University, UK \and 
    Max Planck Institute for Software Systems, Germany
}
\date{}

\begin{document}

\maketitle

\begin{abstract}
Word equations are a crucial element in the theoretical foundation of 
constraint solving over strings, which have received a lot of attention in
recent years. 
A word equation relates two words over string variables and constants. 
Its solution amounts to a function mapping variables to constant strings that 
equate the left and right hand sides of the equation. 
While the problem of solving word equations is decidable, 
the decidability of the problem of solving a word equation
with a length constraint (i.e., a constraint relating the lengths of 
words in the word equation) has remained a long-standing open problem.
In this paper, we focus on the subclass of quadratic word equations, i.e., 
in which each variable occurs at most twice. 
We first show that the length abstractions of solutions to quadratic
word equations are in general not Presburger-definable.
We then describe a class of counter systems with Presburger transition relations
which capture the length abstraction of a quadratic word equation with regular
constraints.
We provide an encoding of the effect of a simple loop of the counter systems
in the theory of existential Presburger Arithmetic with divisibility (PAD).
Since PAD is decidable, we get a decision procedure for quadratic words equations with
length constraints for which the associated counter system is \emph{flat} (i.e.,
all nodes belong to at most one cycle).
We show a decidability result (in fact, also an NP algorithm with a PAD oracle) 
for a recently proposed NP-complete fragment of word equations called regular-oriented word equations,
together with length constraints. 
Decidability holds when the constraints are additionally extended with regular constraints 
with a 1-weak control structure. 
\end{abstract}

\section{Introduction}
\label{sec:intro}

Reasoning about strings is a fundamental problem in computer science and mathematics.
The full first order theory over strings and concatenation is undecidable.
A seminal result by Makanin \cite{Makanin} (see also \cite{Diekert,Jez})
shows that the satisfiability problem for the \emph{existential fragment} is decidable,
by showing an algorithm to check satisfiability of \emph{word equations}. 
Precisely, a word equation $L = R$ consists of two words $L$ and $R$ over an alphabet of 
constants and variables.
Such an equation is satisfiable if there is a mapping $\sigma$ from the variables to strings over the
constants such that $\sigma(L)$ and $\sigma(R)$ are syntactically identical.
\OMIT{
Over the years, Makanin's decidability result has been improved in different ways,
to find alternate, simpler, algorithms, to characterize its complexity, 
and to generalize the result to broader classes, for example, in the presence of
regular constraints on the solutions \cite{Schulz}.
It is known today that the problem is in PSPACE and NP-hard \cite{Diekert,Plandowski,Jez}.
}

An original motivation for studying word equations was to show undecidability of Hilbert's 10th problem (see, e.g., \cite{Matiyasevich68}).
While Makanin's later result shows that word equations could not, by themselves, show undecidability, Matiyasevich in 1968
considered an extension of word equations with \emph{length constraints} as a possible route to showing undecidability of Hilbert's
10th problem \cite{Matiyasevich68}.
A length constraint constrains the solution of a word equation by requiring a linear relationship to
hold on the lengths of words in a solution $\sigma$.
For example, a length constraint might require that a solution maps variable $x$ and variable $y$ to words of the same length. 
The decidability of word equations with length constraints remains open. 

In recent years, reasoning about strings with length constraints has found renewed interest through applications
in program verification and reasoning about security vulnerabilities.
The focus of most research has been on developing practical string solvers 
\cite{fmsd14,S3,BTV09,Abdulla14,cvc4,Balzarotti08,FL10,FPBL13,HAMPI,Z3-str,yu2011,CMS03,popl18-efficient,fang-yu-circuits}.
These solvers are sound but make no claims of completeness.
Relatively few results are known about the decidability status of strings with length and other constraints
(see \cite{Ganesh-boundary} for an overview of the results in this area). 
The main idea in most existing decidability results is the encoding of length constraints into Presburger arithmetic \cite{Ganesh13,Ganesh-boundary}.
However, the length abstraction of a word equation, that is, the set of possible lengths of variables in its solutions,
need not be Presburger definable.
(Indeed, this was Matiyasevich's motivation in studying this problem as a way to prove undecidability of Hilbert's 10th problem.)

In this paper, we consider the case of \emph{quadratic} word equations, in which each variable can appear at most twice \cite{Lentin,diekert-quadratic1},
together with length and regularity constraints.
For quadratic word equations, there is a simpler decision procedure (called the Nielsen transform or Levi's method) based
on a non-deterministic proof tree construction.
The technique can be extended to handle regular constraints \cite{diekert-quadratic1}.
However, we show that already for this class (even for a simple equation like
$xaby = yabx$, where $x,y$ are variables and $a,b$ are constants), the length 
abstraction need not be Presburger-definable.
Thus, techniques based on Presburger encodings are not sufficient to prove decidability.

Our first observation in this paper is a connection between the problem of quadratic word equations with length constraints
and a class of counter systems with Presburger transitions.
Informally, the counter system has control states corresponding to the nodes of the proof tree constructed by Levi's method,
and a counter standing for the length each word variable.
Each step of Levi's method may decrease at most one counter.
Thus, from any initial state, the counter system terminates.
We show that the set of initial counter values which can lead to a successful leaf (i.e., one containing the trivial equation $\epsilon = \epsilon$)
is precisely the length abstraction of the word equation.

Our second observation is that 
the reachability relation for a simple loop of the counter system can be encoded in 
the existential theory of Presburger arithmetic with divisibility $\PAD$.
The encoding is non-trivial in the presence of regular constraints, and depends on structural results on semilinear sets.
As $\PAD$ is decidable \cite{Lipschitz,LOW15}, we obtain a technique
to symbolically represent the reachability relation for \emph{flat} counter systems,
in which each node belongs to at most one loop.

Moreover, the same encoding shows decidability for word equations with length
constraints, provided the proof tree is associated with flat counter systems.
In particular, we show that the class of \emph{regular-oriented} word equations, introduced by \cite{DMN17},
have flat proof trees.
Thus, the satisfiability problem for quadratic regular-oriented word equations with length constraints 
is decidable (and in NEXP).\footnote{
	In fact, the decision procedure is NP with an oracle access to a decision procedure for Presburger arithmetic
	with decidability. The best complexity bounds for the latter are NEXP and NP-hardness \cite{LOW15}.
}

While our decidability result is for a simple subclass, this class is already non-trivial without length and regular
constraints: satisfiability of regular-oriented word equations is NP-complete \cite{DMN17}.
Our result generalizes previous decidability results \cite{Ganesh-boundary}.
Moreover, we believe that the techniques introduced in this paper, such as the connection between acceleration and
word equations, and the use of existential Presburger with divisibility, can 
open the way to more sophisticated decision
procedures or tools based on acceleration designed for counter systems.

\section{Preliminaries}
\label{sec:prelim}

\noindent
\textbf{General notation:}
Let $\N = \Z_{\geq 0}$ be the set of all natural numbers. For integers $i \leq 
j$, we use $[i,j]$ to denote the set $\{i,i+1,\ldots,j-1,j\}$ of integers.
If $i \in \N$, let $[i]$ denote $[0,i]$. We use $\preceq$ to denote the
component-wise ordering on $\N^k$, i.e., $(x_1,\ldots,x_k) \preceq
(y_1,\ldots,y_k)$ iff $x_i \leq y_i$ for all $i \in [1,k]$. If $\bar x \preceq
\bar y$ and $\bar x \neq \bar y$, we write $\bar x \prec \bar y$.

If $S$ is a set,
we use $S^*$ to denote the set of all finite sequences $\gamma = s_1\ldots s_n$ 
over $S$. The length $|\gamma|$ of $\gamma$ is $n$.
The empty sequence is denoted by $\epsilon$. Notice that $S^*$
forms a monoid with the concatenation operator $\cdot$. If $\gamma'$ is a prefix
of $\gamma$, we write $\gamma' \preceq \gamma$. Additionally, if $\gamma' \neq
\gamma$ (i.e. a strict prefix of $\gamma$), we write $\gamma' \prec \gamma$.
Note that the operator $\preceq$ is overloaded here, but the meaning should be
clear from the context.

\smallskip
\noindent
\textbf{Words and Automata:} 
We assume basic familiarity with word combinatorics and automata theory.
Fix a (finite) alphabet $\ialphabet$. 
For each finite word $w := w_1\ldots w_n \in
\ialphabet^*$, we write $w[i,j]$, where $1 \leq i \leq j \leq n$, to denote the segment
$w_i\ldots w_j$. 
We write $\epsilon$ for the empty word.

Two words $x$ and $y$ are \emph{conjugates} if there exist words $u$ and $v$ such that
$x = uv$ and $y = v u$.
Equivalently, $x = \mathrm{cyc}^k(y)$ for some $k$ and for the cyclic
permutation􏲄operation $\mathrm{cyc} : \ialphabet^* \rightarrow \ialphabet^*$, defined as
$\mathrm{cyc}(\epsilon) = \epsilon$, and $\mathrm{cyc}(a\cdot w) = w\cdot a$ 
for $a\in \ialphabet$ and $w \in \ialphabet^*$.

Given a \defn{nondeterministic finite automaton (NFA)} 
$\Aut := (\ialphabet,\controls,\transrel,q_0,q_F)$,
a \defn{run} of $\Aut$ on $w$ is a function $\AutRun: \N
\to \controls$ with $\AutRun(0) = q_0$ that obeys the transition relation 
$\transrel$.
We may also denote the run $\rho$ by the word $\rho(0)\cdots \rho(n)$ over
the alphabet $\controls$. 
The run $\AutRun$ is said to be \defn{accepting} if $\rho(n) = q_F$, in which
case we say that the word $w$ is \defn{accepted} by $\Aut$. The language
$\Lang(\Aut)$ of $\Aut$ is the set of words in $\ialphabet^*$ accepted by
$\Aut$. In the sequel, for $p,q \in \controls$ we will write $\Aut_{p,q}$ to
denote the NFA $\Aut$ with initial state replaced by $p$ and
final is replaced by $q$.

\smallskip
\noindent
\textbf{Word equations:}
Let $\ialphabet$ be a (finite) alphabet of constants and $\Vars$ a set of 
variables; we assume $\ialphabet \cap \Vars = \emptyset$.
A \emph{word equation} $E$ is an expression of the form $L=R$, where $L,R \in
(\ialphabet\cup\Vars)^* \times (\ialphabet\cup\Vars)^*$.
A system of word equations is a nonempty set 
$\set{L_1 = R_1, L_2 = R_2,\ldots, L_k = R_k}$ of word equations.
The length of a system of word equations is the length 
$\sum_{i=1}^k (|L_i|+|R_i|)$.
A system is called \emph{quadratic} if each variable occurs at most twice in all.
A \emph{solution} to a system of word equations 
is a homomorphism $\soln : (\ialphabet\cup\Vars)^* \rightarrow \ialphabet^*$
which maps each $a\in \ialphabet$ to itself that equates the l.h.s. and
r.h.s. of each equation, i.e.,
$\soln(L_i) = \soln(R_i)$ for each $i=1,\ldots,k$.

For each variable $x\in\Vars$, we shall use $|x|$ to denote a formal variable 
that stands for the length of variable $x$.
Let $L_{\Vars}$ be the set $\set{|x|\mid x\in\Vars}$.
A \emph{length constraint} is a formula in Presburger arithmetic whose free 
variables are in $L_\Vars$.

A \emph{solution} to a system of word equations with a length constraint $\Phi(l_{x_1}, \ldots, l_{x_n})$
is a homomorphism $\soln : (\ialphabet\cup\Vars)^* \rightarrow \ialphabet^*$
which maps each $a\in \ialphabet$ to itself such that $\soln(L_i) = \soln(R_i)$ for each $i=1,\ldots,k$
and moreover $\Phi(|\soln(x_1)|, \ldots, |\soln(x_n)|)$ holds.
That is, the homomorphism maps each variable to a word in $\ialphabet^*$ such that each word equation
is satisfied, and the lengths of these words satisfy the length constraint.

The \defn{satisfiability problem} for word equations with length constraints 
asks, 
given a system of word equations and a length constraint, whether it has a solution.

We also consider the extension of the problem with regular constraints.
For a system of word equations, a 
variable $x\in\Vars$, and a regular language $\mathcal{L} \subseteq \ialphabet^*$, 
a \emph{regular constraint} $x\in \mathcal{L}$ imposes the additional restriction
that any solution $\soln$ must satisfy $\soln(x)\in \mathcal{L}$.
Given a system of word equations, a length constraint, and a set of regular constraints,
the satisfiability problem asks if there is a solution satisfying the word equation,
the length constraints, as well as the regular constraints.

In the sequel, for clarity of exposition,
we restrict our discussion to a system consisting of a single word equation (w.l.o.g.).

\smallskip
\noindent
\textbf{Linear arithmetic with divisibility:}
Let $\PA$ be a first-order language with equality, with
binary relation symbol $\leq$, and with terms being linear
polynomials with integer coefficients. 
We write $f(x)$, $g(x)$, etc., for terms in 
integer variables $x = x_1, \ldots , x_n$. 
Atomic formulas in Presburger arithmetic 
have the form $f(x) \leq g(x)$ or
$f(x) = g(x)$.
%
The language $\PAD$ of \emph{Presburger arithmetic with divisibility}
extends the language $\PA$ with a binary relation $\divides$ (for divides).
An atomic formula has the form 
have the form $f(x) \leq g(x)$ or $f(x) = g(x)$ of $f(x) \divides g(x)$,
where $f(x)$ and $g(x)$ are linear polynomials with integer coefficients.
The full first order theory of $\PAD$ is undecidable, but the existential
fragment is decidable \cite{Lipschitz,LOW15}.

Note that the divisibility predicate $x \divides y$ is \emph{not} expressible
in Presburger arithmetic: a simple way to see this is that $\set{(x,y) \in\N^2 \mid x \divides y}$
is not a semi-linear set.

\smallskip
\noindent
\textbf{Counter systems:}
In this paper, we specifically use the term ``counter systems'' to mean
counter systems with Presburger transition relations
(e.g.~see \cite{FAST}). These more general transition relations can be 
simulated by standard Minsky's counter machines, but they are more useful for
coming up with decidable subclasses of counter systems.
A \defn{counter system} $\CA$ is a tuple $(\counters,\controls,\transrel)$, 
where $\counters = \{\ctr_1,\ldots,\ctr_m\}$ is a finite set of counters, 
$\controls$ is a finite set of control states, and $\transrel$ is a finite
set of transitions of the form $(q,\Phi(\bar\ctr,\bar\ctr'),q')$, where
$q,q' \in \controls$ and $\Phi$ is a Presburger formula with free variables
$\ctr_1,\ldots,\ctr_m,\ctr_1',\ldots,\ctr_m'$. A \defn{configuration} of
$\CA$ is a tuple $(q,\vecV) \in \controls \times \N^m$. 

The semantics of counter systems is given as a transition system.
A \defn{transition system} is a tuple $\struct := \transys$,
where $\transysDom$ is a set of \defn{configurations}
and $\to\ \subseteq S \times S$ is a binary relation over $S$.
A \defn{path} in $\struct$ is a sequence $s_0 \to \cdots \to s_n$ of
configurations $s_0,...,s_n \in S$. 

A counter system
$\CA$ generates the transition system $\Trans{\CA} = \transys$, where 
$\transysDom$ is the set of all configurations of $\CA$, and 
$(q,\vecV) \to (q',\vecV')$ if there exists a transition
$(q,\Phi(\bar\ctr,\bar\ctr'),q') \in \transrel$ such that
$\Phi(\vecV,\vecV')$ is satisfiable. 

In the sequel, we will be needing the notion of flat counter systems
\cite{BIL09,LS06,BFLS05,FAST}. 
Given a counter system $\CA = (\counters,\controls,\transrel)$, the
\defn{control structure} of $\CA$ is an edge-labeled directed
graph $G = (V,E)$ with the set $V = \controls$ of nodes and the set 
$E = \transrel$.
The counter system $\CA$ is \defn{flat} if each node $v \in V$ is contained
in at most one simple cycle.

\section{Solving Quadratic Word Equations}
\label{sec:nielsen}

We start by recalling a simple textbook recipe (called Nielsen transformation, 
a.k.a., Levi's Method) \cite{Diekert,Lentin} for solving quadratic word 
equations, both 
for the cases with and without regular constraints. We then discuss the length
abstractions of solutions to quadratic word equations, and provide several
natural examples that are not Presburger-definable.

\subsection{Nielsen transformation}
We will define a rewriting relation $E \To E'$ between quadratic word equations
$E, E'$.
Let $E$ be an equation of the form $\alpha w_1 = \beta w_2$ with 
$w_1,w_2 \in (\ialphabet \cup \Vars)^*$ and $\alpha,\beta \in \ialphabet \cup 
\Vars$. Then, there are several possible $E'$:
\begin{itemize}
    \item \defn{Rules for erasing an empty prefix variable}. These rules can be 
        applied 
        if $\alpha \in \Vars$ (symmetrically, $\beta \in \Vars$). In this case, 
        we can nondeterministically guess that $\alpha$ be the empty 
        word $\epsilon$. That is, $E'$ is $(w_1 = w_2)[\epsilon/\alpha]$.
       The symmetric case of $\beta \in \Vars$ is similar.
    \item \defn{Rules for removing a nonempty prefix}. These rules are
        applicable if each of $\alpha$ and $\beta$ is either a constant or
        a variable that we nondeterministically guess to be a nonempty word.
        There are several cases:
        \begin{description}
            \item[(P1)] $\alpha = \beta$ (syntactic equality). In this 
                case, $E'$ is $w_1 = w_2$.
            \item[(P2)] $\alpha \in \ialphabet$ and $\beta \in \Vars$. In this case,
                $E'$ is $(w_1 = \beta w_2)[\alpha\beta/\beta]$. 
            \item[(P3)] $\alpha \in \Vars$ and $\beta \in \ialphabet$. In this case,
                $E'$ is $(\alpha w_1 = w_2)[\beta\alpha/\alpha]$. 
            \item[(P4)] $\alpha,\beta \in \Vars$. In this case, we
                nondeterministically guess if $\alpha \preceq \beta$ or
                $\beta \preceq \alpha$. In the former case,
                the equation $E'$ is $(w_1 = \beta w_2)[\alpha\beta/\beta]$.
                In the latter case, the equation $E'$ is
                $E'$ is $(\alpha w_1 = w_2)[\beta\alpha/\alpha]$. 
        \end{description}
\end{itemize}
Note that the transformation keeps an equation quadratic. 

\begin{proposition}
    $E$ is solvable iff
    $E \To^* (\epsilon = \epsilon)$. Furthermore,
    checking if $E$ is solvable is in PSPACE.
\end{proposition}

See \cite{Diekert} for a proof. Roughly speaking, the proof uses the fact that each
step either decreases the size of the equation, or the length of a 
length-minimal solution. It runs in PSPACE (in fact, linear space) because each rewriting does not
increase the size of the equation.


\subsection{Handling regular constraints}
Nielsen transformation easily extends to quadratic word equations with regular
constraints (e.g. see \cite{diekert-quadratic1}). We assume that a regular
constraint $x \in \mathcal{L}$ is given as an NFA $\Aut_{p,q}$ representing $\mathcal{L}$. 
[If $q_0$ and $q_F$ are the initial and final states (respectively) of an
NFA $\Aut$, we can be more explicit and write $\Aut_{q_0,q_F}$ instead of 
$\Aut$.] 

Our rewriting relation $\To$ now works over a pair consisting of an equation
$E$ and a set $S$ of regular constraints over variables in $E$. 
Let $E$ be an equation of the form $\alpha w_1 = \beta w_2$ with 
$w_1,w_2 \in (\ialphabet \cup \Vars)^*$ and $\alpha,\beta \in \ialphabet \cup 
\Vars$. We now define $(E,S) \To (E',S')$ by extending the definition of $\To$ 
without regular constraints. In particular, it has to be the case that
$E \To E'$ and additionally do the following:
\begin{itemize}
    \item \defn{Rules for erasing an empty prefix variable $x$}. When applied,
        ensure that each regular constraint $x \in \mathcal{L}$ in $S$ satisfies $\epsilon
        \in \mathcal{L}$. Define $S'$ as $S$ minus all regular constraints of the
        form $x \in \mathcal{L}$.
    \item  \defn{Rules for removing a nonempty prefix}. For (P1), we have
        set $S'$ to be $S$ minus all constraints of the form $\alpha \in \mathcal{L}$ if
        $\alpha$ is a variable. For (P2)--(P4), assume that $E'$ is $(w_1 = 
        \beta
        w_2)[\alpha\beta/\beta]$; the other case is symmetric. For each regular
        constraint $\beta \in \mathcal{L}(\Aut_{p,q})$, we nondeterministically guess a
        state $r$, and add $\alpha \in \mathcal{L}(\Aut_{p,r})$ and $\beta \in
        \mathcal{L}(\Aut_{r,q})$ to $S'$. In the case when $\alpha \in \ialphabet$, we
        could immediately perform the check $\alpha \in \mathcal{L}(\Aut_{p,r})$: a
        positive outcome implies removing this constraint from $S'$, while 
        on a negative outcome our algorithm simply fails on this branch.
        For any variable $y$ that is distinct from $x$,
        we add all regular constraints $y \in L$ in $S$ to $S'$.
\end{itemize}

\begin{proposition}
    $(E,S)$ is solvable iff $(E,S) \To^* (\epsilon = \epsilon,\emptyset)$. 
    Furthermore, checking if $(E,S)$ is solvable is in PSPACE.
\end{proposition}
Note that this is still a PSPACE algorithm because it never creates a new
NFA or adds new states to existing NFA in the regular constraints, but rather 
adds a regular constraint $x \in \mathcal{L}(\Aut_{p,q})$ to a variable $x$, where $\Aut$
is an NFA that is already in the regular constraint.

\subsection{Generating all solutions using Nielsen transformation}
One result that we will need in this paper is that Nielsen transformation
is able to \emph{generate all solutions} of quadratic word equations with 
regular constraints. To clarify this, we extend the definition of $\To$ so that
each a configuration $E$ or $(E,S)$ in the graph of $\To$ is also annotated by
an assignment $\soln$ of the variables to concrete strings. We write 
$E_1[\soln_1] \To E_2[\soln_2]$ if $E_1 \To E_2$ and $\soln_2$ is the
modification from $\soln_1$ according to the operation used to obtain $E_2$ 
from $E_1$. That is, suppose that $\soln_1(x) = ab$ and $\soln_1(y) = abaaab$ 
and $E_1 := xy = yx$ and $E_2 := E_1[xy/y]$. In this case, $\soln_2(x) =
\soln_1(x) = ab$ but $\soln_2(y) = aaab$ since we have taken off the prefix
$\soln_1(x)$ from $\soln_1(y)$. This definition for the case with regular 
constraints is identical.

\begin{proposition}
    $(E,S)[\soln] \to^* (\epsilon = \epsilon,\emptyset)[\soln']$ where 
    $\soln'$ has the empty domain iff
    $\soln$ is a solution of $(E,S)$.
    \label{prop:all-soln}
\end{proposition}
This proposition immediately follows from the proof of correctness of Nielsen
transformation for quadratic word equations \cite{Diekert}.

\subsection{Length abstractions and semilinearity}
Given a quadratic word equation $E$ with constants $\ialphabet$ and variables
$\Vars = \{x_1,\ldots,x_k\}$, its \emph{length abstraction} is defined as 
follows
\[
    \LEN(E) = \{ (|\soln(x_1)|,\ldots,|\soln(x_k)|) : \text{$\soln$ is a
        solution to $E$} \},
    \]
namely the set of tuples of numbers corresponding to lengths of solutions to 
$E$.  
\begin{example}
    Consider the quadratic equation $E:=\ xaby = yz$, where $\Vars = \{x,y,z\}$
    and $\ialphabet$ contains at least two letters $a$ and $b$. We will show
    that its length abstraction $\LEN(E)$ can be captured by the Presburger 
    formula $|x| = |y|+2$. Observe that each $(n_x,n_y,n_z) \in \LEN(E)$
    must satisfy $n_x = n_y + 2$ by a length argument on $E$. Conversely,
    we will show that each triple $(n_x,n_y,n_z) \in \N^3$ satisfying
    $n_x = n_y + 2$ must be in $\LEN(E)$. To this end, we will define a solution
    $\soln$ to $E$ such that $(|\soln(x)|,|\soln(y)|,|\soln(z)|) =
    (n_x,n_y,n_z)$. Consider $\soln(x) = a^{n_x}$. 
    Let $n_z = q(n_x+2)+r$ for
    some $q \in \N$ and $r \in [n_x+1]$. Let $w$ be a prefix of $\soln(x)$ of 
    length $r$. Define $\soln(z) = \soln(x)^qw$. Therefore, $\soln(z)$ is a 
    prefix of $\soln(x)ab\soln(z)$. That is, for some $v \in \ialphabet^*$,
    it is the case that $\soln(z)v = \soln(x)ab\soln(z)$. Letting 
    $\soln(y) = v$ suffices to make $\soln$ satisfy $E$. Also, $|\soln(y)| =
    |\soln(x)ab\soln(z)| - |\soln(z)| = |\soln(x)| + 2$, which satisfies
    the formula $|x|=|y|+2$. \qed
\end{example}
However, it turns out that Presburger Arithmetic is not sufficient for
capturing length abstractions of quadratic word equations.
\begin{theorem}
There is a quadratic word equation whose length abstraction is not
    Presburger-definable.
\end{theorem}
To this end, we show that the length abstraction of $xaby = yabx$, where 
$a,b \in \ialphabet$ and $x,y \in \Vars$, is not Presburger definable.
\begin{lemma}
    The length abstraction $\LEN(xaby = yabx)$ coincides with tuples 
    $(|x|,|y|)$ of numbers satisfying the expression $\varphi(|x|,|y|)$
    defined as:
    \begin{eqnarray*}
        |x| = |y| & \vee & (|x| = 0 \wedge |y| \equiv 0 \pmod{2}) 
                  \vee (|y| = 0 \wedge |x| \equiv 0 \pmod{2}) \\
                  & \vee & (|x|,|y| > 0 \wedge \gcd(|x|+2,|y|+2) > 1)
    \end{eqnarray*}
\end{lemma}
Observe that this would imply non-Presburger-definability: for otherwise, since 
the first three disjuncts are Presburger-definable, the last disjunct would also
be Presburger-definable, which is not the case since the property that two 
numbers are relatively prime is not Presburger-definable.
Note however, that the expression is definable in existential Presburger arithmetic with
divisibility.

Let us prove this lemma. Let $S = \LEN(xaby = yabx)$. We first show that given
any numbers $n_x, n_y$ satisfying $\varphi(n_x,n_y)$, there are
solutions $\soln$ to $xaby = yabx$ with $\soln(\alpha) = n_\alpha$ for each
$\alpha \in \{x,y\}$. If they satisfy the first disjunct in $\varphi$ (i.e.
$n_x = n_y$), then set $\soln(x) = \soln(y)$ to an arbitrary word $w \in
\ialphabet^{n_x}$. If they satisfy the second disjunct, then $aby = yab$
and so set $\soln(x) = \epsilon$ $\soln(y) \in (ab)^*$. The same goes with the
third disjunct. For the fourth disjunct (assuming the first three disjuncts
are false), let $d = \gcd(n_x+2,n_y+2)$. 
Define $\soln(x), \soln(y) \in (a^{d-1}b)^*(a^{d-2})$ so that
$|\soln(\alpha)| = n_\alpha$ for $\alpha \in \Vars$. It follows that
$\soln(x)ab\soln(y) = \soln(y)ab\soln(x)$.

We now prove the converse. So, we are given a solution $\soln$ to $xaby = yabx$
and let $u := \soln(x)$, $v := \soln(y)$. Assume to the contrary that 
$\varphi(|u|,|v|)$ is false and that $u$ and $v$ are the shortest such 
solutions. We have several cases to consider:
\begin{itemize}
    \item $u = v$. Then, $|u| = |v|$, contradicting that $\varphi(|u|,|v|)$
        is false.
    \item $u = \epsilon$. Then, $abv = vab$ and so $v \in (ab)^*$, which implies
        that $|v| \equiv 0 \pmod{2}$. Contradicting that $\varphi(|u|,|v|)$
        is false.
    \item $v = \epsilon$. Same as previous item and that $|u| \equiv 0
        \pmod{2}$.
    \item $|u| > |v| > 0$. Since $\varphi(|u|,|v|)$ is false, we have
        $\gcd(|u|+2,|v|+2) = 1$.
        It cannot be the case that $|u| = |v|+1$ since then,
        comparing prefixes of $uabv = vabu$, the letter at position
        $|u|+2$ would be $b$ on l.h.s. and $a$ on r.h.s., which is a
        contradiction. Therefore $|u| \geq |v|+2$. 
        Let $u' = u[|v|+3,|u|]$, i.e., $u$ but with its prefix of length
        $|v|+2$ removed. By Nielsen transformation, we have
        $u'abv = vabu'$. It cannot be the case that $u' = \epsilon$; for,
        otherwise, $abv = vab$ implies $v\in (ab)^*$ and so $u = vab$, implying
        that 2 divides both $|u|+2$ and $|v|+2$, contradicting 
        that $\gcd(|u|+2,|v|+2) = 1$. Therefore, $|u'| > 0$. Since
        $\gcd(|u'|+2,|v|+2) = \gcd(|u|+2,|v|+2) = 1$, we have a shorter
        solution to $xaby = yabx$, contradicting minimality.
    \item $|v| > |u| > 0$. Same as previous item.
\end{itemize}
%
\OMIT{
we assume that $|\soln(x)| > 1$ (otherwise, it is
trivially true).
Next we can show by induction on $n \in \N$ that if 
$n|\soln(x)| \leq |\soln(z)| < (n+1)|\soln(x)|$, then $|\soln(z)| =
n|\soln(x)|$. The base case is when $n = 0$, which is trivial. So, assume that
$n > 0$. In this case, $\soln(z) = \soln(x)u$ for some $u

let us assume that $|\soln(x)| > 1$ and
$|\soln(z)| > 0$ (otherwise, it
is trivially true). Then, since $\soln(x)[1] = \#$ and $\soln(x)[2] \in (a+b)$,
matching the leftmost letter on both sides of $\soln(xz) = 
\soln(zy)$, we have $\soln(z) \geq 2$, $\soln(z)[1] = \#$ and 
$\soln(z)[2] \in (a+b)$. Since $\soln(y)[1] = \#$ and $\soln(x)[1,|\soln(x)|] 
\in (a+b)^*$, it must be the case that $|\soln(z)| \geq |\soln(x)|$. 
$|\soln(x) = \

We will then
}

\section{Reduction to Counter Systems}
\label{sec:reduce}
\OMIT{
Given a quadratic word equation $E$ with constants $\ialphabet$ and variables
$\Vars = \{x_1,\ldots,x_k\}$, its \emph{length abstraction} is defined as 
follows
\[
    \LEN(E) = \{ (|\soln(x_1)|,\ldots,|\soln(x_k)|) : \text{$\soln$ is a
        solution to $E$} \},
    \]
namely the set of tuples of numbers corresponding to lengths of solutions to 
$E$. 
}
In this section, we will provide an algorithm for computing a counter system
from $(E,S)$, where $E$ is a quadratic word equation and $S$ is a set of regular
constraints. We will first describe this algorithm for the case without
regular constraints, after which we show the extension to the case with
regular constraints.

Given the quadratic word equation $E$, we show how to compute a counter 
system $\CA(E) = (\counters,\controls,\transrel)$ such that the following 
theorem holds.
\begin{theorem}
    The length abstraction of $E$ coincides with 
    $pre_{\CA(E)}^*(\{\epsilon = \epsilon\} \times \N^{|\Vars|})$.
    \label{th:reduce}
\end{theorem}
Before defining $\CA(E)$, we define some notation.
Define the following formulas:
\begin{itemize}
\item $\ID(\bar x,\bar x') := \bigwedge_{x \in \bar x} x' = x$
\item $\SUB_{y,z}(\bar x,\bar x') := z \leq y \wedge y' = y - z \wedge
        \bigwedge_{x \in \bar x, x \neq y} x' = x$
\item $\DEC_{y}(\bar x,\bar x') := y > 0 \wedge y' = y - 1 \wedge 
        \bigwedge_{x \in \bar x, x \neq y} x' = x$
\end{itemize}
Note that the $\neq$ symbol in the guard of $\bigwedge$ denotes syntactic
equality (i.e. not equality in Preburger Arithmetic). We omit mention of the
free variables $\bar x$ and $\bar x'$ when they are clear from the context.

We now define the counter system. 
Given a quadratic word equation $E$ with constants $\ialphabet$ and variables
$\Vars$, we define a counter system 
$\CA(E) = (\counters,\controls,\transrel)$ as follows. 
The counters $\counters$ will be precisely all variables that appear in $E$, i.e., $\counters := \Vars$.
The control states are precisely all equations $E'$ that can be rewritten
from $E$ using Nielsen transformation, i.e., $\controls := \{ E' : E \To^* E'
\}$. The set $\controls$ is finite (at most exponential in $|E|$) as per our 
discussion in the previous section. 

We now define the transition relation $\transrel$. 
We use $\bar x$ to enumerate $\Vars$ in some order.
Given $E_1 \To E_2$ with $E_1,E_2 \in \controls$, we then
add the transition $(E_1,\Phi(\bar x,\bar x'),E_2)$, where $\Phi$ is defined as
follows:
\begin{itemize}
    \item If $E_1 \To E_2$ applies a rule for erasing an empty prefix variable
        $y \in \bar x$, then $\Phi := y = 0 \wedge \ID$.
    \item If $E_1 \To E_2$ applies a rule for removing a nonempty prefix:
        \begin{itemize}
            \item If (P1) is applied, then $\Phi = \ID$.
            \item If (P2) is applied, then $\Phi = \DEC_{\beta}$.
            \item If (P3) is applied, then $\Phi = \DEC_{\alpha}$.
            \item If (P4) is applied and $\alpha \preceq \beta$, then $\Phi = 
                \SUB_{\beta,\alpha}$. If $\beta \preceq \alpha$, then
                $\Phi = \SUB_{\alpha,\beta}$.
        \end{itemize}
\end{itemize}
Observe that if $(E_1,\vecV_1) \to (E_2,\vecV_2)$, then $|E_1| \leq |E_2|$ and
$\vecV_1 \preceq \vecV_2$. In addition, if $\vecV_1 = \vecV_2$, then $|E_1| 
< |E_2|$. This implies the following lemma.
\begin{lemma}
    The counter system $\CA(E)$ terminates from every configuration
    $(E_0,\vecV_0)$.
\end{lemma}
The proof of Theorem \ref{th:reduce} immediately follows from Proposition
\ref{prop:all-soln} that Nielsen transformation generates all solutions. 

\subsubsection*{Extension to the case with regular constraints: } 
\OMIT{
As for Nielsen transformation with to quadratic word equations with regular 
constraints, the extension of our reduction is only needed for applications of 
rules (P2)--(P4). 
}
In this extension, we will only need to assert that the counter values
belong to the length abstractions of the regular constraints, which
are effectively semilinear due to Parikh's Theorem \cite{Parikh}. 
Given a quadratic word equation $E$ with a set $S$ of regular constraints,
we define the counter system $\CA(E,S) = (\counters,\controls,\transrel)$
as follows.
Let $\CA(E) = (\counters_1,\controls_1,\transrel_1)$. 
Let $\counters = \counters_1$. 
Let $\controls$ be the finite set of all configurations reachable from 
$(E,S)$, i.e., $\controls = \{ (E',S') : (E,S) \To^* (E',S') \}$. Given 
$(E_1,S_1) \To (E_2,S_2)$, we add the transition $((E_1,S_2),\Phi(\bar x,\bar
x'),(E_2,S_2))$ as follows. Suppose that $(E_1,\Phi'(\bar x,\bar x'),E_2)$
was added to $\transrel_1$ by $E_1 \To E_2$. Then,
\[
    \Phi' := \Phi \wedge \bigwedge_{x \in \bar x} \left( x \in 
             \LEN(\bigcap_{(x \in L) \in S} L) \wedge
                x' \in \LEN(\bigcap_{(x \in L) \in S'} L)\right).
            \]
The size of the NFA for $\bigcap_{(x \in L)\in S} L$ is exponential in
the number of constraints of the form $(x \in L)$ in $S$ (of which there are
polynomially many). 
The constraint $x \in \LEN(L)$ is well-known to be effectively semilinear 
\cite{Parikh}, and in fact we can compute using the algorithm of
Chrobak-Martinez \cite{Chrobak,Martinez,IPL09}
in polynomial time two finite sets $A, A'$ of integers and an integer $b$
such that,
for each $n \in \N$, $n \in U := A \cup (A'+b\N)$ is true iff 
$n \in \LEN(L)$. Note that $U$ is a fintie union of arithmetic progressions
(with period 0 and/or $b$).
In fact, each number $a \in A \cup A'$ (resp.~the number $b$)
is at most quadratic
in the size of the NFA, and so it is a polynomial
\footnote{Note that we mean polynomial in the size of the NFA, which
can be exponential in $|S|$.} size even when they are written in unary. 
Therefore, treating $U$ as an existential Presburger formula
$\varphi(x)$ with one free variable (an existential quantifier is needed
to guess the coefficient $n$ such that $x = a_i+bn$ for some $i$),
the resulting $\Phi'$ is a polynomial-sized existential Presburger formula.

\begin{theorem}
    The length abstraction of $(E,S)$ coincides with 
    $pre_{\CA(E,S)}^*(\{(\epsilon = \epsilon,\emptyset)\} \times \N^{|\Vars|})$.
    \label{th:reduce2}
\end{theorem}
As for the case without regular constraints, the proof of Theorem 
\ref{th:reduce} immediately follows from Proposition
\ref{prop:all-soln} that Nielsen transformation generates all solutions. 

\OMIT{
\begin{itemize}
    \item If $E_1 \To E_2$ applies a rule for erasing an empty variable or 
        (P1), then $\Phi := \Phi \wedge \bigwedge_{(x \in L) \in S} x \in 
        \LEN(L)$.
    \item Suppose $E_1 \To E_2$ applies (P2)--(P4). Then, assuming $E_1 :=
        \alpha w_1 = \beta w_2$ and $E_2 := E_1[\alpha\beta/\beta]$, we
        define 
\end{itemize}
}

\section{Decidability via Linear Arithmetic with Divisibility}
\label{sec:decidability}

\subsection{Accelerating a 1-variable-reducing cycle}
Consider a counter system $\CA = (\counters,\controls,\transrel)$ with
$\counters = \{q_0,\ldots,q_{n-1}\}$, and for some $y \in \counters$
the transition relation $\transrel$ consists of precisely the following
transition $(q_i,\Phi_i,q_{i+1 \pmod{n}})$, for each $i \in 
[n-1]$, such that $\Phi_i$ is either $\SUB_{y,z}$ (with $z$ a variable distinct
from $y$) or $\DEC_y$. Such a counter system is said to be a 
\defn{1-variable-reducing cycle}. 

\begin{lemma}
    There exists a polynomial-time algorithm which given a 1-variable-reducing
    cycle $\CA = (\counters,\controls,\transrel)$ and two states $p,q \in
    \controls$ computes an formula $\varphi_{p,q}(\bar x,\bar x')$ in 
    existential Presburger with divisibility such that $(p,\vecV) \to_{\CA}^*
    (q,\vecW)$ iff $\varphi_{p,q}(\vecV,\vecW)$ is satisfiable.
    \label{lm:accelerate}
\end{lemma}
This lemma can be seen as a special case of the acceleration lemma for flat 
parametric counter automata \cite{BIL09} (where all variables other than $y$ are
treated as parameters). However, its proof is in fact quite simple.
Without loss of generality, we assume that $q = q_0$ and $p = q_i$, for some $i
\in \N$.
Any path $(q_0,\vecV) \to_{\CA}^* (q_i,\vecW)$ can be decomposed into 
the cycle $(q_0,\vecV) \to^* (q_0,\vecV')$ and the simple path $(q_0,\vecW_0)
\to \cdots \to (q_i,\vecW_i)$ of length $i$. Therefore, the reachability
relation $(q_0,\vecX) \to_{\CA}^* (q_i,\vecY)$ can be expressed as
\[
    \exists \vecZ_0,\cdots,\vecZ_{i-1}: \varphi_{q_0,q_0}(\vecX,\vecZ_0) \wedge 
                \Phi_0(\vecZ_0,\vecZ_1) \wedge \cdots \wedge
                \Phi_{i-1}(\vecZ_{i-1},\vecY).
            \]
Thus, it suffices to show that $\varphi_{q_0,q_0}(\vecX,\vecX')$ is expressible
in $\PAD$. Define a multiset $M$ of counter decrements as follows:
\begin{itemize}
    \item The number of the integer constant 1 $M$ can contain is defined as
        the number of $i$ such that $\Phi_i = \DEC_y$.
    \item For each variable $x \in \Vars\setminus\{y\}$, the number of times
        $x$ could appear in $M$ is defined as the number of $i$ such that
        $\Phi_i = \SUB_{y,x}$.
\end{itemize}
For any variable/constant $e$, we will write $M(e)$ to denote the number of
times $e$ appears in $M$. Therefore, for some $n \in \N$ we have 
$y' = y - n\sum_{e}M(e)$, or 
equivalently
\[
    n\sum_{e}M(e) = y - y'
\]
The formula $\varphi_{q_0,q_0}$ can be defined as
follows:
\[
    \varphi_{q_0,q_0} :=\ \left(\sum_{e}M(e)\right) \mid (y-y') \wedge 
                        y' \leq y \wedge
                            \bigwedge_{x \in \Vars\setminus\{y\}} x' = x.
\]

\smallskip
\noindent
\textbf{Handling unary Presburger guards: }
Recalling our reduction for the case with regular constraints from Section
\ref{sec:reduce} reveals that we also need unary Presburger guards on the
counters. We will show how to extend our aforementioned acceleration lemma to
handle such guards. As we will see shortly, we will need a bit of the theory of
semilinear sets.

As before, our counter system $\CA = (\counters,\controls,\transrel)$ has
counters $\counters = \{q_0,\ldots,q_{n-1}\}$, and the control structure is
a simple cycle of length $n$, i.e., the transitions in $\transrel$ are precisely
$(q_i,\Phi_i,q_{i+1 \pmod{n}})$ for some Presburger formula
$\Phi_i(\bar x,\bar x')$, for each $i \in [n-1]$. We say that $\CA$ is
\defn{1-variable-reducing with unary Presburger guards} if there exists
a counter $y \in \counters$ such that each $\Phi_i$
is of the form $\theta_i \wedge \psi_i$, where $\theta_i$ is either
$\SUB_{y,z}$ (with $z$ a variable distinct from $y$) or $\DEC_y$,
and $\psi_i$ is a conjunction of formulas of the form $x \in A_i \cup 
(A_i'+b\N)$, where both $A_i$ and $A_i'$ are finite sets of natural numbers
and $x \in \counters$.
For each counter $x \in \counters$, we use $\psi_{i,x}$ to denote the set of
conjuncts in $\psi_i$ that refers to the counter $x$.

\begin{lemma}
    There exists a polynomial-time algorithm which given a 1-variable-reducing
    cycle with unary Presburger guards $\CA = (\counters,\controls,\transrel)$ 
    and two states $p,q \in
    \controls$ computes an formula $\lambda_{p,q}(\bar x,\bar x')$ in 
    existential Presburger with divisibility such that $(p,\vecV) \to_{\CA}^*
    (q,\vecW)$ iff $\lambda_{p,q}(\vecV,\vecW)$ is satisfiable.
    \label{lm:accelerate2}
\end{lemma}
Unlike Lemma \ref{lm:accelerate}, this lemma does not immediately follow
from the results of \cite{BIL09} on flat parametric counter automata.
To prove this, let us first take the formula $\varphi_{p,q}(\bar x,\bar x')$
from Lemma  \ref{lm:accelerate} applied to $\CA'$, which is obtained from
$\CA$ by first removing the unary Presburger guards. We can insert these unary
Presburger guards to $\varphi_{p,q}$, but this is not enough because we need to
make sure that all ``intermediate'' values of $y$ have to also satisfy the
Presburger guards corresponding to $y$ on that control state. More precisely,
let the counter decrement in $\theta_i$ be $\alpha_i$ (which can either be
a variable $x$ distinct from $y$ or 1). For $j \in [n-1]$, we use $f_j(\bar x)$ 
to denote $\sum_{i=0}^{j} \alpha_i$. Write $f(\bar x)$ for $f_{n-1}(\bar x)$.
Then, we can write
    \begin{eqnarray*}
        \lambda_{q_0,q_0} & := & \bar x' = \bar x \vee \left( 
\varphi_{q_0,q_0} \wedge \bigwedge_{i=0}^{n-1} \psi_i(\bar x) \wedge \eta_{q_0,q_0} \right) \\
        \eta_{q_0,q_0}    & := & \forall k: 
                           y' + (k+1)f(\bar x) \leq y \longrightarrow \\
                          &    & \quad \left(\bigwedge_{i=0}^{n-1}
            \bigwedge_{(\alpha_i \in A\cup A'+b\N) \in \psi_{i,y}}
                y' + kf(\bar x) + \alpha_i \in A \cup (A'+b\N)
            \right)
    \end{eqnarray*}
This is a correct expression that captures the reachability relation
$(q_0,\vecW) \to_{\CA}^* (q_0,\vecW')$, but the problem is that it has a
universal quantifier and therefore is not a formula of $\PAD$. To fix this
problem, we will need to exploit the semilinear structure of unary Presburger
guards. To this end, we first notice that, by taking the big conjunction over 
$i$ and the big conjuncton over $\alpha_i$ out, the formula $\eta_{q_0,q_0}$ is 
equivalent to:
\begin{eqnarray*}
    \eta_{q_0,q_0}    & \equiv & \bigwedge_{i=0}^{n-1} 
                    \bigwedge_{(\alpha_i \in A\cup A'+b\N) \in \psi_{i,y}}
                    \forall k: 
                           y' + (k+1)f(\bar x) \leq y \longrightarrow \\
                          &    & \quad (
                y' + kf(\bar x) + \alpha_i \in A \cup (A'+b\N))
\end{eqnarray*}
Therefore, it suffices to rewrite each conjunct 
$C(\bar x) := \forall k: y' + (k+1)f(\bar x) 
\leq y \longrightarrow \quad ( y' + kf(\bar x) + \alpha_i \in A \cup (A'+b\N)$ 
as an existential Presburger formula, for each $i$ and constraint 
$(\alpha_i \in A\cup A'+b\N)$. To this end, let $a := \max A$ and let $N$
denote $|A'|$. We claim that 
\begin{eqnarray*}
    C(\bar x) & \equiv & \bigwedge_{i=0}^a y' + if(\bar x) + \alpha_i\leq y
                \to y' + if(\bar x) + \alpha_i \in A\cup (A'+b\N) \\
              &        & \wedge \bigwedge_{i=a+1}^{a+N+1} 
                                y' + if(\bar x) + \alpha_i\leq y
                \to y' + if(\bar x) + \alpha_i \in A'+b\N. \\
\end{eqnarray*}
Simply put, we distinguish the cases when $y' + if(\bar x) + \alpha_i$ is
``small'' (i.e., less than the maximum threshold that can keep this number
in an arithmetic progression with 0 period), and when this number is
``big'' (i.e. must be in an arithmetic progression with a nonzero period). 
To prove this equivalence, it suffices to show that if $y' + kf(\bar x) +
\alpha_i \notin A\cup (A'+b\N)$ with $k > a+N+1$ and
$y' + kf(\bar x) + \alpha_i \leq y$, then we can find $k' \leq a+N+1$ such that
$y' + k'f(\bar x) + \alpha_i \notin A\cup (A'+b\N)$. Suppose to the contrary
that such $k'$ does not exist. Then, since there are $N+1$ numbers in between
$a+1$ and $a+N+1$, by pigeonhole principle there is an arithmetic progression 
$a' + b\N$ and two different numbers $a+ 1 \leq j_1 < j_2 \leq a+N+1$ such that
$y' + j_h f(\bar x) + \alpha_i \in a'+b\N$, for $h=1,2$. Let $d := (j_2-
j_1)$. Note that $df(\bar x)$ denotes the difference between $y' + j_1 f(\bar x)
+ \alpha_i$ and $y' + j_2 f(\bar x) + \alpha_i$, and this difference
is of the form $m b$, for some positive integer $m$. We now find a number
$j \in [a+1,a+N]$ with $j+qd = k$ for some positive integer $q$. Since $y' + j
f(\bar x) + \alpha_i \in a''+b\N$ for some $a'' \in A'$, it must be the case 
that $y' + (j+qd) f(\bar x)+ \alpha_i \in a''+b\N$ for $q \in \N$, contradicting
that $y' + kf(\bar x) + \alpha_i \notin A\cup(A'+b\N)$.

We have proven correctness, and what remains is to analyse the size of the
formula $\lambda_{q_0,q_0}$. To this end, it suffices to show that each formula
$C(\bar x)$ is of polynomial size. This is in fact the case since there are 
at most polynomially many numbers in $A$ and $A'$ and that the size of all
numbers in $A \cup A' \cup \{b\}$ are of polynomial size even when they are written
in unary.

\subsection{An extension to flat control structures and an acceleration scheme}

The following generalisation to flat control structures is an easy corollary of 
Lemma \ref{lm:accelerate} and \ref{lm:accelerate2}.

\begin{theorem}
    There exists a polynomial-time algorithm which, given a flat Presburger
    counter system $\CA = (\counters,\controls,\transrel)$, each of whose
    simple cycle is 1-variable-reducing with unary Presburger guards
    and two states $p,q \in
    \controls$, computes an formula $\lambda_{p,q}(\bar x,\bar x')$ in 
    existential Presburger with divisibility such that $(p,\vecV) \to_{\CA}^*
    (q,\vecW)$ iff $\lambda_{p,q}(\vecV,\vecW)$ is satisfiable.
    \label{th:flat}
\end{theorem}
Indeed, to prove this theorem, we can simply use Lemma \ref{lm:accelerate2}
to accelerate all cycles and the fact that transition relations expressed in 
existential Presburger with divisibility is closed under composition.

\OMIT{
Theorem \ref{th:flat} gives rise to a straightforward \emph{general loop 
acceleration scheme $\ACCELERATE$} 
for underapproximating the solutions to a quadratic word equation $(E,S)$ with
regular constraints. First, we build the counter system $\CA(E,S) =
(\counters,\controls,\transrel)$. Next, a \defn{flattening} of $\CA(E,S)$ is
a flat counter system $\CA' = (\counters,\controls',\transrel')$, each of
whose simple cycle is 1-variable-reducing with unary Presburger guards,
such that there exists a mapping $g: \controls' \to \controls$ (called
\defn{folding}) such that: (1) $(p,\Phi,q) \in \transrel'$ implies
$(g(p),\Phi,g(q)) \in \transrel$, (2) there exists $p,q \in \controls'$
such that $g(p) = (E,S)$ and $g(q) = (\epsilon=\epsilon,\emptyset)$. Therefore,
$\ACCELERATE$ enumerates all flattenings $\CA_1,\CA_2,\ldots$ of $\CA(E,S)$ in 
some order, and 

Next, given
any positive integer $k$, we define the $k$-unfolding of $\CA(E,S)$ as the
counter system $\CA^k(E,S) = (\counters,\controls_k,\transrel_k)$, where:
\begin{itemize}
    \item $\controls_k$ is the set of all paths in the control structures of
        $\CA(E,S)$ of length at most $k$.
    \item 
\end{itemize}
}

\subsection{Application to word equations with length constraints}
Theorem \ref{th:flat} gives rise to a simple and sound (but not complete)
technique for solving quadratic word equations with length constraints:
given a quadratic word equation $(E,S)$ with regular constraints, if
the counter system $\CA(E,S)$ is flat, each of whose simple cycle is
1-variable-reducing with unary Presburger guards, then apply the decision
procedure from Theorem \ref{th:flat}. In this section, we show completeness of
this method for the class of regular-oriented word equations recently defined 
in \cite{DMN17}, which can be extended with regular constraints given as 1-weak
NFA \cite{BRS12}.
\OMIT{
A word $w\in (\ialphabet\cup\Vars)^*$ is \emph{regular-(ordered)} if each 
variable $x\in\Vars$ appears at most once
in $w$.
The word $w\in (\ialphabet\cup\Vars)^*$ is \emph{non-crossing} if between any two occurrences of the same variable $x$,
no other variable different from $x$ occurs.
For example $ax_1ax_2$ is regular but $ax_1bx_1$ is not, and $ax_1bx_1x_2ax_2$ is non-crossing but 
$ax_1bx_2x_1ax_2$ is not.
}
A word equation is \defn{regular} if each variable $x
\in\Vars$ occurs at most once on each side of the equation. 
Observe that $xy = yx$ is regular, but $xxyy = zz$ is not. It is easy to
see that a regular word equation is quadratic.
A word equation $L = R$ is said to be 
\defn{oriented} if there is a total ordering $<$ on $\Vars$ such that 
the occurrences of variables on each side of the equation preserve $<$, i.e.,
if $w = L$ or $w = R$ and $w = w_1\alpha w_2\beta w_3$ for some $w_1,w_2,w_3 \in
(\ialphabet\cup\Vars)^*$ and $\alpha,\beta \in \Vars$, then $\alpha < \beta$.
Observe that $xy = yz$ (i.e. that $x$ and $z$ are conjugates) is oriented, but
$xy = yx$ is not oriented. It was shown in \cite{DMN17} that the satisfiability
for regular-oriented word equations is NP-hard.
We show satisfiability for this class with length constraints is decidable.

\begin{theorem}
    The satisfiability problem of regular-oriented word equations with
    length constraints is decidable in nondeterministic exponential time.
    \label{th:sat}
\end{theorem}
This decidability (in fact, an NP upper bound) for the
\emph{strictly regular-ordered} subcase (i.e. each variable occurs precisely
once on each side) was proven in \cite{Ganesh-boundary}. 
For solving this subcase, it was shown that Presburger Arithmetic is sufficient,
but the decidability for the general class of regular-oriented word equations 
with length constraints remains open, which we prove in this paper. 
\begin{lemma}
Given a regular-oriented word equation $E$, the counter system
    $\CA(E)$ is flat. Moreover, the length of each simple 
    cycle (resp.~path) in the control structure of $\CA(E)$ is of length 
    $O(|E|)$ (resp.~$O(|E|^2)$).
    \label{lm:rowq_is_flat}
\end{lemma}
Before proving this lemma, we show a simple lemma that $\To$ preserves
regular-orientedness. Its proof can be found in the appendix.
\begin{lemma}
    If $E \To E'$ and $E$ is regular-oriented, then $E'$ is also
    regular-oriented.
    \label{lm:preserve}
\end{lemma}
We now prove Lemma \ref{lm:rowq_is_flat}.
\OMIT{
\begin{lemma}
    Given a simple cycle $E_0 \To E_1 \To \cdots \To E_n$ with $n > 0$ (i.e. 
    $E_0 = E_n$ and $E_i \neq E_j$ for all $0 \leq i < j < n$). Suppose
    that $E_i := L_i = R_i$ with $L_i = \alpha_i w_i$ and $R_i = \beta_i w_i'$. 
\end{lemma}
}
Let $E := L = R$. We first show that the length of a simple cycle in the
control structure of $\CA(E)$ 
is of length at most $N = \max\{|L|,|R|\} - 1$. Given a simple cycle
$E_0 \To E_1 \To \cdots \To E_n$ with $n > 0$ (i.e. $E_0 = E_n$ and $E_i \neq 
E_j$ for all
$0 \leq i < j < n$), it has to be the case that each rewriting in this cycle
applies one of the (P2)--(P4) rules since the other rules reduce the size of the
equation. We have $|E_0| = |E_1| = \cdots = |E_n|$.
Let $E_i := L_i = R_i$ with $L_i = \alpha_i w_i$ and $R_i = \beta_i w_i'$. 
Let us assume that $E_1 = E_0[\alpha_0\beta_0/\beta_0]$; the case with
$E_1 = E_0[\beta_0\alpha_0/\alpha_0]$ will be easily seen to be symmetric. 
This assumption implies that $\beta_0$ is a variable $y$, and that 
$L_0 = uyv$ for some words $u,v \in (\ialphabet \cup \Vars)^*$ (for, otherwise,
$|E_1| < |E_0|$).
Furthermore, 
it follows that, for each $i \in [n-1]$, $E_{i+1} =
E_i[\beta_i\alpha_i/\beta_i]$, i.e., the counter system $\CA(E)$ applies
either $\SUB_{y,x}$ (in the case when $x = \alpha_i$) or $\DEC_y$ (in the
case when $\alpha_i \in \ialphabet$). For, otherwise, taking a minimal $i \in
[1,n-1]$ with $E_i[\beta_i\alpha_i/\alpha_i]$ for some variable $x = \alpha_i$
shows that $E_i$ is of the form $x ... y ... = y ... x ...$ (since $|E_{i+1}| =
|E_i|$) contradicting that $E_i$ is oriented. Consequently, we have
\begin{itemize}
    \item $R_i = R_j$ for all $i,j$, and 
    \item $L_i = \mathrm{cyc}^i(u)yv$ for all $i \in [n]$
\end{itemize}
implying that the length of the cycle is at most $|R_0| - 1 \leq |R| - 1$.

Consider the control structure $\CA(E)$ as a dag of SCCs. In this dag,
each edge from one SCC to the next is size-reducing. 
Therefore, the maximal length of a path in 
this dag is $|E|$. Therefore, since the maximal path of each SCC is $N$ (from
the above analysis), the maximal length of a simple path in the control
structure is at most $N^2$.

\smallskip
\noindent
\textbf{Handling regular constraints: } It is difficult to extend 
Theorem \ref{th:sat} to the case with regular constraints because 
they may introduce nestings of cycles (which breaks the flat control
structure) even for regular-oriented word equation. However, we can show that
restricting to regular constraints given by \defn{1-weak NFA} \cite{BRS12}
(i.e. a dag of SCCs, each with at most one state) preserves the flat control
structure. The class of 1-weak automata
is in fact quite powerful, e.g., when considered as recognisers of
languages of $\omega$-words, they capture the subclass of LTL with operators 
$\Fu$ and
$\Gl$ \cite{BRS12}. They have also been used to obtain a decidable extension of
infinite-state concurrent systems in term rewriting systems, e.g., see
\cite{KRS09,TL10}. In the context of quadratic word equations,
we can use 1-weak NFA to capture the regular constraint $x,y \in \#(a+b)^*$, 
which in conjunction with $xy = yz$ gives rise a non-Presburger length
abstraction. Such an NFA will have two states $q_0$ and $q_1$, and transitions
$q_0 \stackrel{\#}{\longrightarrow} q_1$ and $q_1 
\stackrel{a,b}{\longrightarrow} q_1$, where $q_0$ is an initial state and 
$q_1$ a final state.

\begin{theorem}
    The satisfiability problem of regular-oriented word equations with
    1-weak regular constraints and length constraints is solvable in 
    nondeterministic double exponential time (2NEXP).
    \label{th:sat_reg}
\end{theorem}
Let us prove this theorem. Suppose $E$ is a regular-oriented word equation
with the set $S$ of 1-weak regular constraints. Let $\CA(E,S) =
(\counters,\controls,\transrel)$ be the corresponding counter system.
Let $M(S)$ denote the maximum number of states ranging over all NFA in $S$.
\begin{lemma}
    The counter system $\CA(E,S)$ is flat. Moreover, the length of each simple 
    cycle in the control structure of $\CA(E,S)$ is of length 
    $O(|E|)$, while the length of each simple path is of length 
    $O(|E|^2|\Var||S|M^3)$.
    \label{lm:rowq_reg_is_flat}
\end{lemma}
By virtue of Theorem \ref{th:flat}, this lemma implies decidability of 
Theorem \ref{th:sat_reg}, but it does NOT imply the nondeterministic exponential
time upper bound since each unary Presburger guard in $\CA(E)$ will be
of the form $x \in \LEN(\bigcap_{(x \in L) \in S} L)$. Even though we know
that $|S|$ is always of a polynomial size, their intersection requires
performing a product automata construction, which will result in an NFA of an
exponential size. Therefore, we obtain a nondeterministic double exponential
time complexity upper bound (2NEXP), instead of NEXP as for the case without
regular constraints.
The proof of Lemma \ref{lm:rowq_reg_is_flat} can be found in the appendix.
\OMIT{
By exploiting the structure of 1-weak automata,
it turns out that $x \in \LEN(\bigcap_{(x \in L) \in S}$
\emph{can} be captured by a polynomial-sized existential Presburger formula,
which would immediately imply the nondeterministic exponential time upper
bound.
\begin{lemma}
    Let $\Aut_1, \ldots, \Aut_m$ be 1-weak NFA over the alphabet $\ialphabet$.
    Then, we can comnpute in polynomial-time an existential Presburger
    formula $\varphi(x)$ that captures $\LEN(\bigcap_{i=1}^m \Lang(\Aut_i))$.
    \label{lm:weak_len}
\end{lemma}
}

\OMIT{
Our algorithm will first nondeterministically guess
a simple path $P$ of SCCs from an initial state $q_0$ to a final state $q_F$ in 
each 1-weak NFA, i.e., of the form $q_0 \stackrel{a_0}{\longrightarrow} q_1 
\stackrel{q_1}{\longrightarrow} \cdots \stackrel{a_{n-1}}{\longrightarrow} q_n$,
where 
$q_n = q_F$, $q_i \neq q_j$ for each $1 \leq i < j \leq n$, and that each
$q_i$ could potentially have a self-loop (i.e. $q_i 
\stackrel{A_i}{\longrightarrow} q_i$ for a set $A_i \subseteq \ialphabet$
of letters in the alphabet).
}

\begin{remark}
    Our proof of Theorem \ref{th:sat_reg} does not extend to the case when we 
    allow \emph{generalised flat} NFA (i.e. after mapping all the letters in
    $\ialphabet$ to a new symbol '?', the control structure of the NFA is 
    flat) in the regular constraints. This is because a simple cycle involving 
    two or more states will result in a counter system that is no longer flat.
\end{remark}

Finally, we mention that the length abstraction of 
regular-oriented word equations with regular constraints is in general not 
Presburger-definable (see appendix for proof).
\begin{proposition}
    The regular-oriented word equation $xy = yz$ with regular constraints
    $x,y \in \#(a+b)^*$ has non-Presburger-definable length abstraction.
    \label{prop:nonPres_reg}
\end{proposition}


\section{Future Work}
\label{sec:conc}

\OMIT{
We have shown that solving quadratic word equations with
regular constraints and length constraints requires us to use a decidable
logic beyond Presburger Arithmetic. The notion of length abstractions of
equational constraints is crucial when we are
additionally given length constraints in the input. We have developed a sound
method for computing the length abstractions quadratic word equations with
regular constraints by exploiting connections to flat counter systems 
and existential Presburger Arithmetic with divisibility $\PAD$.
Our method is complete for the class of regular-oriented word 
equations with regular constraints given as 1-weak NFA. In the case
without (resp.~with) regular constraints, our algorithm runs in
nondeterministic polynomial (resp.~exponential) time using $\PAD$ solvers 
as a blackbox. Since the complexity of $\PAD$ is a major open problem and is
known to be in between NP and NEXP \cite{LOW15}, we have obtained a NEXP
(resp.~2NEXP) complexity upper bound for the problem. An improvement on the
complexity of $\PAD$ will result in a better complexity for our problem.
}
One obvious research direction is to study extensions of 
our techniques to deal with the class of regular (but not necessarily oriented)
word equations with length constraints. We believe that this is a key 
subproblem of the general class of quadratic word equations with length
constraints. Finally, we conjecture that the length abstractions of general
quadratic word equations with regular constraints can be effectively
captured by existential Presburger with divisibility.

\paragraph{Acknowledgment.} We thank Artur Je\.{z} for the constructive feedback
and his lecture notes on word equations. We thank Dmitry Chistikov, Matthew
Hague, Philipp R\"{u}mmer, and James Worrell for the helpful discussions.


\shortlong{}{
\appendix

\clearpage
\begin{center}
    \LARGE \bfseries APPENDIX
\end{center}

\section{Proof of Lemma \ref{lm:preserve}}
    It is easy to see each rewriting rule preserves regularity. Now, because
    $E'$ is regular, to show that $E' := L = R$ is also oriented it is
    sufficient and necessary to 
    show that there are no two variables $x, y$ such that $x$ occurs before
    $y$ in $L$, but $y$ occurs before $x$ in $R$. All rewriting rules except 
    for (P2)--(P4) are easily seen to preserve orientedness. Let us write
    $E := \alpha w_1 = \beta w_2$ with $\alpha \neq \beta$, and assume $E' = 
    E[\alpha\beta/\beta]$; the
    case of $E' = E[\beta\alpha/\alpha]$ is symmetric. So, $\beta$ is some 
    variable $y$. If $\beta$ does not occur in $w_1$, then $L = 
    w_1$ and $R = \beta w_2$ and that $E$ is oriented
    implies that $E'$ is oriented. So assume that $\beta$ appears in $w_1$,
    say, $w_1 = u\beta v$. Then, $R= \beta w_2$ and $L = u\alpha\beta v$.
    Thus, if $\alpha \in \ialphabet$, $E'$ is oriented because we can use
    the same variable ordering that witnesses that $E$ is oriented. So, assume
    $\alpha \in \Vars$. It suffices to show that $\alpha$ occurs 
    \emph{at most once} in $E$.
    For, if $\alpha$ also occurs on the other side of the equation $E$
    (i.e. in $w_2$), $\alpha$ precedes $\beta$ on l.h.s. of $E$, while
    $\beta$ precedes $\alpha$ on r.h.s. of $E$, which would show that $E$
    is not oriented. 
%

\section{Proof of Lemma \ref{lm:rowq_reg_is_flat}}
    Since $\CA(E,S)$ is simply $\CA(E)$ but annotated by the regular 
    constraints and how they are modified, it suffices to prove that each 
    SCC $G$ in the control structure of $\CA(E)$ with nodes $E_1,\ldots,E_m$
    the restrictions $\CA'$ of $\CA(E,S)$ to control states of the form
    $(E_i,S)$, for some set $S$ of regular constraints, is a flat each of whose
    cycle is of length $O(|E|)$, and each of whose simple path is of
    length at most $O(|E||\Var||S|M^3)$. By Theorem \ref{th:sat}, we know that
    $\CA(E)$ is flat, each of whose simple cycle is 1-variable-reducing and
    is of length $|E|$ and each of whose simple path over the dag of SCCs
    corresponding to the control structure of $\CA(E)$ is of length at most
    $O(|E|)$. Therefore, we may assume that the SCC $G$ is a cycle
    $E_1 \to \cdots \to E_m$ and there exists a counter $y$ which is reduced
    by each transition in $G$ using $\SUB_{y,z}$ or $\DEC_y$. 

    We will next define a partial order $\geqReg$ on sets of regular constraints
    of the form $x \in L(\Aut_{p,q})$, where $x \in \Vars$ and $\Aut$ is an NFA 
    in $S$. We will write $S_1 \gtReg S_2$ if $S_1 \geqReg S_2$ but $S_1 \neq 
    S_2$. Before defining $\geqReg$,
    the intuition behind this partial order is that each transition 
    $((E',S'),\Phi,(E'',S''))$ in the control structure $G'$ of $\CA'$ will 
    $\geqReg$-decrease $S'$, i.e.,
    $S' \geqReg S''$. In particular, this implies that every
    simple cycle in $G'$ will consist of the nodes $(E_1,H),
    \ldots,(E_m,H)$ for some set $H$ of regular constraints. Therefore,
    the length of each simple cycle in $G'$ is at most $O(|E|)$.
    Since we will
    see that the length a simple path in $\geqReg$ is at most $O(|\Var||S|M^3)$,
    it will follow that the length of a simple path in $G'$ is at most
    $O(|E||\Var||S|M^3)$.

    We write $S_1 \geqReg S_2$ if: 
    \begin{itemize}
        \item The number of regular constraints in $S_2$ on $y$ (i.e. of the 
            form $y \in L$ for some $L$) is at most the number of
            regular constraints in $S_1$ on $y$.
        \item For each variable $x$ distinct from $y$, if $x \in L$ is in
            $S_1$, then it is also in $S_2$.
        \item For each constraint $y \in \Lang(\Aut_{r,q})$ in $S_2$, there
            exists a state $p$ that can reach $r$ in $\Aut$ such that the 
            constraint $y \in \Lang(\Aut_{p,q})$ is in $S_1$.
    \end{itemize}
    The fact that $\geqReg$ is a partial order then follows from the fact
    that each $\Aut$ is 1-weak, i.e., whose transition relation gives rise
    to a partial order on the states of $\Aut$. To show that the length of
    a simple path in $\geqReg$ is at most $O(|\Var||S|M^3)$, observe that
    we can add at most $|S|M^2$ regular constraints on $x$ (distinct from
    $y$). In addition, if the constraints on $y$ in some $S_1$
    are 
    \[
        y \in \Lang(\Aut_{p_1,q_1}^1), \ldots, y \in \Lang(\Aut_{p_K,q_K}^K),
    \]
    and if $k_i$ is the length of maximal simple path in $\Aut_{p_i,q_i}^i$,
    then one can change the set of constraints on $y$ in $S_1$ to any
    $S_2 \leqReg S_1$ by at most $\sum_{i=1}^K k_i = O(|S|M^3)$. This proves
    that any simple path in $\geqReg$ is of length $O(|\Var||S|M^3)$.

\section{Proof of Proposition \ref{prop:nonPres_reg}}
We claim that its length abstraction is
precisely the set of triples $(n_x,n_y,n_z) \in \N^3$ satisfying the formula
\[
    \varphi(l_x,l_y,l_z) := l_x = l_y \wedge l_x > 0 \wedge l_x \mid l_z.
\]
Since divisibility is not Presburger-definable,
the theorem immediately follows. 
To show that for each triple $\bar n = (n_x,n_y,n_z)$ satisfying $\varphi$
there exists a solution $\soln$ to $E$ and the constraint
$x,y \in \#(a+b)^*$, simply consider $\soln$ with 
$\soln(x) = \soln(y) = \#a^{l_x - 1}$, and
$\soln(z) = \soln(z) = \soln(x)^{n_z/n_x}$. Conversely, consider a solution
$\soln$ satisfying $xz = zy$ and $x,y \in \#(a+b)^*$. We must have $x = y$
since two conjugates $x,y \in \#(a+b)^*$ must apply a full cyclical permutation,
i.e., the same words. 
We then have
$|\soln(x)| = |\soln(y)| > 0$. To show that
$|\soln(x)| \mid |\soln(z)|$, let $|\soln(z)|= q|\soln(x)| + r$ for some
$q \in \N$ and $r \in [|\soln(x)|-1]$. It suffices to show that $r = 0$.
To this end, matching both sides of $E$, we obtain $z = x^qw$, where $w$ is a
prefix of $\soln(x)$ of length $r$. If $r > 0$, then matching both sides of the
equation from the \emph{right} reveals that the last $|\soln(y)|-1$ letters
on l.h.s. of $\soln(E)$ contains $\#$, which is not the case on r.h.s. of
$\soln(E)$, contradicting that $\soln$ is a solution to $E$. Therefore, $r = 0$,
proving the claim.
}

\end{document}